\documentclass[apjl]{emulateapj}
\usepackage{amssymb,multirow,color,natbib}

\renewcommand{\vec}[1]{\mathbf{#1}}
\renewcommand{\(}{\left(}
\renewcommand{\)}{\right)}

\begin{document}
\title{Scaling properties of small-scale fluctuations in magnetohydrodynamic turbulence}
\author{Jean Carlos Perez$^{1}$, Joanne Mason$^2$, Stanislav Boldyrev$^{3,4}$, Fausto Cattaneo$^5$}
\affiliation{${~}^1$Space Science Center, University of New Hampshire, Durham, NH 03824 \\ 
  ${~}^2$College of Engineering, Mathematics and Physical Sciences, University of Exeter, EX4 4QF, UK\\
  ${~}^3$Department of Physics, University of Wisconsin at Madison, 1150 University Ave, Madison, WI 53706
  \\ ${~}^4$Kavli Institute for Theoretical Physics, University of  California at Santa Barbara, Santa Barbara, CA 93106
  \\ ${~}^5$Department of Astronomy and Astrophysics, University of  Chicago, 5640 S. Ellis Ave, Chicago, IL 60637\\ 
  {\sf jeanc.perez@unh.edu}, {\sf j.mason@exeter.ac.uk}, {\sf    boldyrev@wisc.edu}, {\sf cattaneo@flash.uchicago.edu} }

\begin{abstract} Magnetohydrodynamic (MHD) turbulence in the majority of natural systems, including the interstellar medium, the solar corona, and the solar wind, has Reynolds numbers far exceeding the Reynolds numbers achievable in numerical experiments. Much attention is therefore drawn to the universal scaling properties of small-scale fluctuations, which can be reliably measured in the simulations and then extrapolated to astrophysical scales. However, in contrast with hydrodynamic turbulence, where the universal structure of the inertial and dissipation intervals is described by the Kolmogorov self-similarity, the scaling for MHD turbulence cannot be established based solely on dimensional arguments due to the presence of an intrinsic velocity scale -- the Alfv\'en velocity. In this Letter, we demonstrate that the Kolmogorov first self-similarity hypothesis cannot be formulated for MHD  turbulence in the same way it is formulated for the hydrodynamic case.  Besides profound consequences for the analytical consideration, this also imposes stringent conditions on numerical studies of MHD  turbulence. In contrast with the hydrodynamic case, the discretization scale in numerical simulations of MHD turbulence should decrease faster than the dissipation scale, in order for the simulations to remain  resolved as the Reynolds number increases.   
\end{abstract}

\maketitle

\section{Introduction}
The energy distribution over scales in magnetic plasma turbulence 
is the important input ingredient in theories of the interstellar medium \citep[][]{elmegreen_s2004,brandenburg_n2011}, scintillation of galactic radio sources \citep[][]{goldreich_s2006,coles_etal2010}, particle heating and acceleration by magnetic plasma fluctuations in the solar wind \citep[][]{chandran_etal10,chandran_dqb2011}. At scales much larger than the plasma micro-scales (such as the ion gyroscale and the ion inertial length) many fundamental 
aspects of the plasma dynamics can be captured in the framework of MHD~\citep[e.g.,][]{biskamp03,terry09,tobias_cb11}, which can be effectively studied both analytically and numerically. 

In spite of advances in present-day computer simulations, the Reynolds numbers of astrophysical flows ($Re\sim 10^6$--$10^{16}$) exceed by many orders of magnitude the Reynolds numbers 
achieved numerically ($Re\sim 10^4$). In this situation, major interest is attracted to the scaling properties of MHD turbulence, which can be reliably established from numerical simulations. Such an approach motivates phenomenological models that can be extrapolated to astrophysically relevant scales. Examples include models of astrophysical dynamo action \citep[e.g.,][]{malyshkin_b2010,brandenburg_etal2012,kolekar_etal2012}, models of magnetic reconnection at high Lundquist numbers in the solar wind and the solar corona  \citep[e.g.,][]{longcope_s1994,rappazzo_etal2008,ng_etal2012,zhdankin_etal2013}, and studies of turbulent mixing in the interstellar medium~\citep[e.g.,][]{sur_etal2014}, etc.

The basic 
assumptions of universality and scale invariance that are common in studies of hydrodynamic turbulence are not well-justified and, as a result, not well-understood in the MHD case, and they require careful investigation. In the hydrodynamic case, the Kolmogorov first self-similarity hypothesis implies that at scales much smaller than the driving scale, the energy spectrum of incompressible non-magnetized fluid turbulence has a universal form \cite[][]{kolmogorov1941,obukhov1941}: 
\begin{eqnarray}
{E(k)}=C_k \epsilon^{2/3}k^{-5/3} g_h(k\eta_h),
\label{espec}
\end{eqnarray}
where $\epsilon$ is the mean energy dissipation rate, 
\begin{eqnarray}
\eta_h=\nu^{3/4}\epsilon^{-1/4} 
\label{eta_h}
\end{eqnarray}
is the Kolmogorov viscous scale, and $\nu$ is the fluid viscosity. The function $g_h(x)$ is expected to be universal, that is, independent of the nature of the large-scale driving, and to satisfy $g_h(0)=1$. With the driving applied at scale $L$, in the inertial interval of turbulence $kL\gg 1 \gg k\eta_h$ the function $g_h(x)\rightarrow 1$ as $\eta_h\rightarrow 0$, thus leading to the well-known $k^{-5/3}$ Kolmogorov's inertial range spectrum.  At the dissipation scales, $k\eta_h\gtrsim 1$,  the form of the function $g_h$ cannot be derived from scaling arguments. However, it has been  constrained by detailed experimental and numerical measurements \cite[e.g.,][]{tsuji2004,donzis2010} and phenomenologically modeled \cite[e.g.,][]{monin_yaglom1975}.  

In MHD turbulence the Kolmogorov self-similarity relation~(\ref{espec}) does not apply 
due to the presence of the Alfv\'en velocity $v_A=B_0/\sqrt{4\pi \rho}$ associated with the large-scale magnetic field $B_0$. The large-scale magnetic field mediates the turbulent dynamics at small scales, therefore, the energy spectrum may essentially depend on the large scale \cite[][]{iroshnikov,kraichnan65,goldreich_s95,ng_b96,galtier_nnp00,bhattacharjee_n01,boldyrev05,boldyrev06}. In this case, the general form of the energy spectrum can be written as 
\begin{eqnarray}
E(k)=C^M_k \epsilon^{2/3} k^{-5/3} g(k\eta, k\Lambda),
\end{eqnarray}
where $\eta $ is the dissipation scale and $\Lambda\sim L$ is the scale parameter related to the large-scale organization of the flow -- both can be different in different setups. The mediation of the small-scale interaction by the large-scale magnetic field implies that in the inertial interval, $k\eta \ll 1$, one cannot require that $g(0, k\Lambda)=1$. Rather, in order to establish the energy spectrum in this case one needs to study the nonlinear interaction of Alfv\'en wave packets in detail \cite[][]{iroshnikov,kraichnan65,goldreich_s95,ng_b96,galtier_nnp00,bhattacharjee_n01,boldyrev05,boldyrev06}. 

\section{Self-similarity in weak and strong MHD turbulence.}
We start with the case of balanced weak MHD turbulence, where the average energies in oppositely propagating Alfv\'en wave packets are the same. We assume a strong uniform background field, $B_0\gg b_{\rm rms}$, and suppose that turbulence is isotropically excited at scale $L$, such that $v_{\rm rms}\sim b_{\rm rms}/\sqrt{4\pi \rho}$. The weakness of the interaction follows from the fact that the linear Alfv\'en frequency, $v_A/L$, is much larger than the frequency of nonlinear interaction, $v_{\rm rms}/L$. 
It has been derived that the inertial-interval energy spectrum of balanced MHD turbulence scales with the field-perpendicular wavenumber as $k_{\perp}^{-2}$ \cite[e.g.,][]{galtier_nnp00,boldyrev_p09,wang11}, which allows us to write the asymptotics of the spectral function in the form
\begin{eqnarray}
g_w(0, k_\perp\Lambda)\sim (k_\perp\Lambda)^{-1/3}.
\label{gweak}
\end{eqnarray}  
One can demonstrate that $\Lambda\sim L$, and $\eta_w=\nu (v_A/\epsilon \Lambda)^{1/2}$. Thus, we observe that the inertial interval essentially depends on the outer scale.

A similar consideration applies in the regime of steadily driven balanced strong MHD turbulence. It has been found in numerical simulations and phenomenological models that the field-perpendicular energy spectrum in this case scales as $k_{\perp}^{-3/2}$  \cite[][]{muller_g05,maron_g01,haugen_04,boldyrev05,boldyrev06,chen10,mason_cb08,perez_etal2012}, which implies the following asymptotic for the spectral function: 
\begin{eqnarray}
g_s(0, k_\perp\Lambda)\sim (k_\perp\Lambda)^{1/6}.
\label{gstrong}
\end{eqnarray} 
Here $\Lambda$ is related to the cross-helicity of the flow (for example, $\Lambda \sim L$, if the magnetic and velocity fluctuations are driven at the outer scale in a non-correlated fashion), and 
\begin{eqnarray}
\eta_s = \nu^{2/3} \Lambda^{1/9} \epsilon^{-2/9}  
\label{eta_s}
\end{eqnarray}
is the dissipation scale, see, e.g.,~\cite[][]{perez_etal2012}. In the
imbalanced case it has also been shown that the asymptotic of the
spectrum follows Equation (\ref{gweak}) in the weak imbalanced
case~\citep[][]{boldyrev_p09} and Equation~(\ref{gstrong}) in the strong
imbalanced case~\citep[][]{perez_b09,podesta_b10}.  

Both examples demonstrate that the dependence on $\Lambda$ in  MHD turbulence is crucial for establishing the energy distribution in both the inertial and dissipation intervals. In both cases the spectrum deviates from the Kolmogorov $k^{-5/3}$ due to a reduction of the nonlinear interaction by a certain mechanism related to the large-scale magnetic field. In the case of weak turbulence, such a mechanism is the decorrelation of the triple-field products due to the short crossing time of counter-propagating Alfv\'en waves \cite[e.g.,][]{ng_b96,galtier_nnp00,bhattacharjee_n01}. In the case of strong turbulence, a weakening of the nonlinear interaction is provided by the scale-dependent angular alignment between magnetic and velocity fluctuations, that is, progressive ``Alfv\'enization" of the turbulence at small scales \cite[e.g.,][]{boldyrev05,boldyrev06,mason_cb08}. As we argue in the next section, the dependence of the spectral function $g_s(k\eta, k\Lambda)$  on the outer scale is crucial for the applicability of discrete numerical schemes for simulations of MHD turbulence. 

\section{The problem of numerical resolution in simulations of MHD turbulence.} 
In this section we concentrate on strong MHD turbulence, and assume that the simulations are performed in a numerical scheme discretized at scale $\Delta$, which can be the grid size of a finite-difference scheme, the inverse dealiasing cut-off of a pseudo-spectral scheme, etc. In the presence of a numerical cutoff $\Delta$, the general form of the function $g_s$ is $g_s(k\eta,k\Lambda, k\Delta)$. The solution of the discrete scheme in general is different from the physical solution  and it may 
have different scaling properties as it contains an additional dimensional parameter $\Delta$. However, it needs to converge  to the physical solution as $\Delta\to 0$. 

It should be recalled that in the hydrodynamic case, when the spectrum is independent of $\Lambda$, the $g$ function can be written as $g_h(k\eta_h, \Delta/\eta_h)$. Therefore, as long as the numerical resolution is a fixed fraction of the dissipation scale $\Delta/\eta_h=\mbox{const}$, the  form of the function $g_h(k\eta_h)$ is universal \cite[e.g.,][]{gotoh2002}. In contrast, the presence of an additional scale $\Lambda$ in the MHD case means that there are infinitely many $g_s$ functions that provide the same inertial interval asymptotics, but have different behavior at the subrange scales. Consider, for example, a spectral function\footnote{This function is given for illustrative purposes and is not chosen to match a particular numerical simulation.} 
\begin{eqnarray}
g_s
=\left[\frac{1}{k\Lambda}+\frac{\Delta}{\sqrt{\eta_s\Lambda}}\right]^{-1/6}g_1(k\eta_s, k\Delta), \,\,\,\,\,
\label{example}
\end{eqnarray}
where $g_1(0,0)=1$. 
The expression (\ref{example}) agrees with the inertial scaling of $-3/2$, while it steepens at small scales due to the finite discretization cutoff~$\Delta$. In this example, the scaling of the numerically measured energy spectrum changes at scale $k\sim  \sqrt{\eta_s/\Lambda}(1/\Delta)$ from $-3/2$ to $-5/3$; however, this spectral steepening represents the property of the numerical scheme, not of the physical solution. In order to avoid the influence of numerical effects, one needs to ensure that the discretization cutoff is sufficiently small. For instance, in order to observe the inertial interval up to the dissipation scale~$k\sim 1/\eta_s$, one needs to require that~$\Delta<\eta_s^{3/2}/\Lambda^{1/2}$. 

In this example the discretization scale in the numerical simulations needs to decrease faster than the dissipation scale, for the  simulations to be resolved. In the next section we demonstrate that 
a similar situation is encountered in simulations of MHD turbulence.

\section{Numerical simulations.}
The incompressible MHD equations can be written in the Els\"asser form:
\begin{eqnarray}
 \label{mhd-elsasser}
  \left( \frac{\partial}{\partial t}\mp\vec v_A\cdot\nabla \right)
  \vec z^\pm+\left(\vec z^\mp\cdot\nabla\right)\vec z^\pm &=& -\nabla
  P +\nu\nabla^2 \vec z^{\pm}+\vec f^\pm, \nonumber \\ \nabla \cdot
  {\vec z}^{\pm}&=&0,
  \end{eqnarray}
where $\vec z^\pm=\vec v\pm\vec b$ are the Els\"asser variables, and $\vec
v$ and $\vec b$ are the fluctuating velocity and 
magnetic fields in units of the Alfv\'en velocity, $\vec{v_A}={\bf
  B}_0/\sqrt{4\pi\rho_0}$. In these equations $P=(p/\rho_0+b^2/2)$, where 
$p$ is the plasma pressure, $\rho_0$ is the background plasma density and
$\nu$ is the fluid viscosity. For simplicity, 
the viscosity 
is equal to the magnetic diffusivity. The turbulence is driven at large scales by the forces~$\vec f^\pm$. In the linear case, the plasma waves 
can be decomposed into shear Alfv\'en waves whose polarizations are perpendicular to both $\vec{B_0}$ and the wave-vector $\vec{k}$, and pseudo-Alfv\'en waves whose polarizations are in the plane of $\vec{B_0}$ and $\vec{k}$ and perpendicular to $\vec{k}$.

In the case of strong MHD turbulence, the pseudo-Alfv\'en modes are dynamically irrelevant for
the turbulent cascade \cite[e.g.,][]{goldreich_s95}. 
One can therefore filter out the pseudo-Alfv\'en modes by setting $\vec z^\pm_\|=0$, which reduces the equations to the Reduced MHD model:
\begin{eqnarray}
  \(\frac{\partial}{\partial t}\mp\vec v_A\cdot\nabla_\|\)\vec
  z^\pm_{\perp}+\left(\vec z^\mp_{\perp}\cdot\nabla_\perp\right)\vec z^\pm_{\perp} =
  -\nabla_\perp P\nonumber\\ +\nu\nabla^2\vec z^\pm_{\perp} +\vec f_\perp^\pm. 
  \label{rmhd-elsasser}
\end{eqnarray}  
We note that in RMHD the fluctuating fields have only two vector
components, but that each depends on all three spatial
coordinates. 
Due to incompressibility, each field has only one degree of freedom, which can be expressed in terms of stream functions in the more standard form of
the RMHD equations~\citep[][]{strauss76}. The equivalence
between RMHD and MHD in the strong turbulence regime has been shown
in numerical simulations; for an extensive discussion
see~\cite{mason_etal12}. In light of this 
equivalence, we will refer to the numerical spectrum obtained from
RMHD simulations as the MHD spectrum. We solve the
RMHD equations (\ref{rmhd-elsasser}) in a periodic, rectangular domain
with aspect ratio $L_{\perp}^2 \times L_\|$, where the subscripts
denote the directions perpendicular and parallel to $\vec{B_0}$,
respectively. We set $L_{\perp}=2\pi$, $L_\|/L_\perp=6$ and
$\vec{B_0}=5\vec{e_z}$.  A fully dealiased three-dimensional pseudo-spectral 
algorithm is used on a grid with a resolution of $N_{\perp}^2\times
N_\|$ mesh points.

Both Els\"asser variables are driven by independent random forces $\vec{f^+}$ and $\vec{f^-}$ applied in Fourier space at wave-numbers $2\pi/L_{\perp} \leq k_{x,y} \leq 2 (2\pi/L_{\perp})$, $k_\| =
2\pi/L_\|$. The forces have no component along $z$ and they are solenoidal
in the $xy$ plane.  Their Fourier coefficients  are Gaussian
random numbers with amplitudes chosen so that $v_{\rm rms}\sim 1$. The
individual random values are refreshed independently on average
approximately $10$ times per turnover time of the large-scale eddies. The
variances $\sigma_{\pm}^2=\langle |\vec f^{\pm} |^2\rangle$ control
the average rates of energy injection into the $z^+$ and $z^-$
fields. In this work we discuss the regime of balanced MHD
turbulence, i.e., $\sigma^+ \approx \sigma^-$. 
The Reynolds number is defined as $Re=v_{\rm rms}(L/2\pi)/\nu$.

\begin{figure}
\includegraphics[width=\columnwidth]{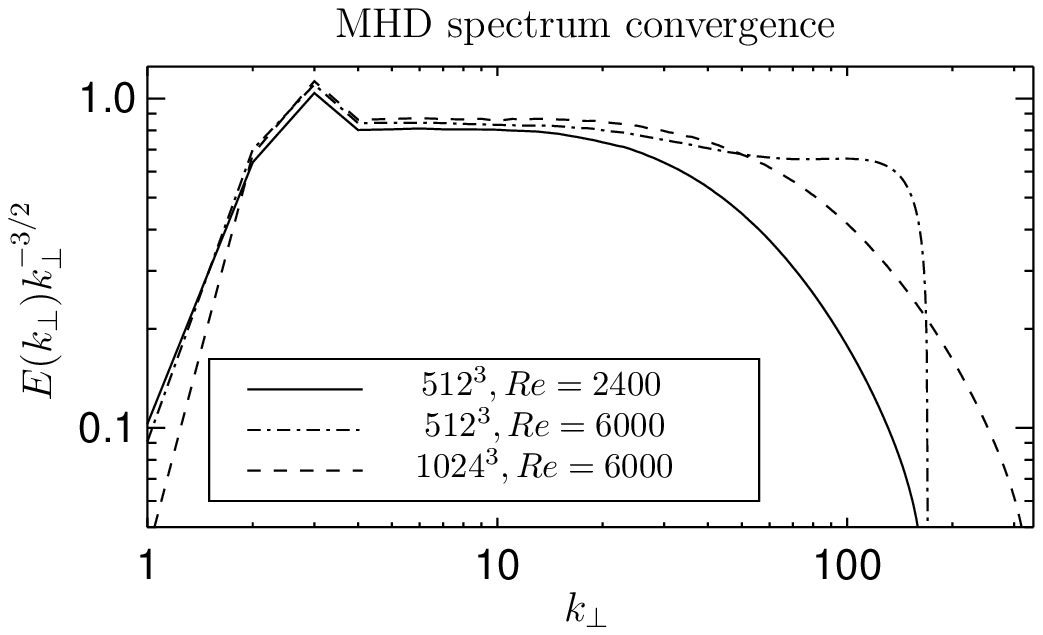}
\caption{\label{fig:spectra1}Resolution study of the numerical spectrum in MHD turbulence. The solid line represents the energy spectrum in numerical simulations of MHD turbulence at $512^3$ collocation points and $Re=2400$. The inertial interval is well-resolved in this case. The dash-dotted line represents a similar run where the Reynolds number is increased to $Re=6000$. The latter simulation is unresolved; as a result, the numerical solution does not approximate the physical one at small scales. The numerical spectrum steepens at $k\ge15$, and then flattens closer to the cut-off scale, which is a purely numerical effect. This numerical effect disappears as the resolution is increased to $1024^3$ without changing the physical parameters of the simulations (the dashed curve),  where the inertial interval now  extends to about $k\sim 25$.}
\end{figure}

\section{The results.} 
The spectra of MHD turbulence obtained in the simulations are shown in Figure.~\ref{fig:spectra1}. The solid line represents the energy spectrum for a $512^3$, $Re=2400$ run, which is well-resolved. The dash-dotted line shows the same set up with decreased viscosity, which makes the dissipation scale approach the numerical cut-off. The scales at $k>15$ are now  significantly affected by the proximity to the dealiasing cut-off \cite[cf.][]{perez_etal2012}. The proximity to the $k$-space cutoff is known to distort the spectral behavior at small scales in hydrodynamic simulations \cite[e.g.,][]{cichowlas_2005,frisch_etal2008,connaughton2009,grappin2010}; our MHD runs bear similarity with those results. Such a spectral distortion is a property of the numerical scheme and it 
should not be confused  with the inertial interval behavior. Indeed, as one increases the number of grid points to $1024^3$ (thereby reducing $\Delta$ to $\Delta/2$), while keeping all the physical parameters unchanged, the numerical distortion disappears, see the dashed line in Figure.~\ref{fig:spectra1}. 

\begin{figure}
\includegraphics[width=\columnwidth]{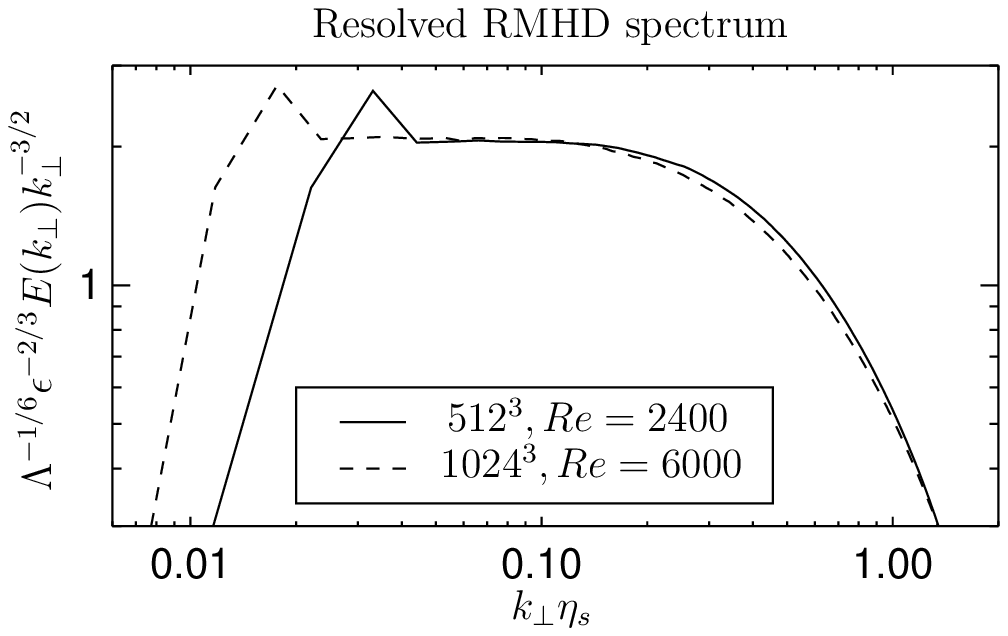}
\caption{\label{fig:eta_s}This figure illustrates the scaling of a fully resolved numerical spectrum.  The solid curve corresponds to a resolution of $512^3$ and $Re=2400$, the dashed curve corresponds to a resolution $1024^3$ and $Re=6000$. In both runs $\Delta/\eta_s\approx 0.17\ll 1$, so that both simulations are resolved at small scales. Here, $\Delta = L_\perp/N_\perp$. As a result, the solution of the numerical scheme is self-similar with $g_s(0,k_\perp\Lambda)$ given by Equation (\ref{gstrong}).} 
\end{figure}

We now note that in the resolved runs, the dissipation interval scales according to the  formula (\ref{eta_s}), see Figure.~\ref{fig:eta_s}, and also \cite[][]{perez_etal2012}. However, in the unresolved runs, the scaling of the small-scale spectrum is different and is consistent with expression~(\ref{eta_h}), as is seen in Figure.~\ref{fig:spectra2}. This is not surprising; as was discussed in the previous section, the scaling of the numerical solution close to the discretization cutoff may be different from the scaling of the physical solution if the former does not approximate the latter. Such a scaling of unresolved runs was previously observed in \cite[][]{beresnyak_11,beresnyak_12,beresnyak_14}, where it was incorrectly attributed to the scaling of the physical solution because the numerical convergence was not checked.

\section{Discussion.}
\begin{figure}
\includegraphics[width=\columnwidth]{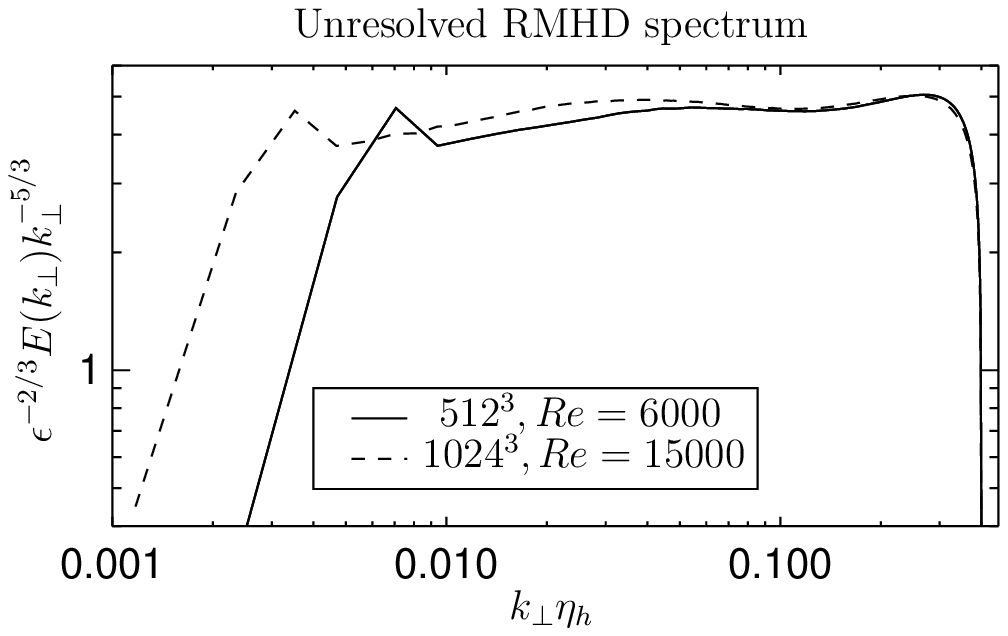}
\caption{\label{fig:spectra2}This figure illustrates the scaling of
  the unresolved numerical spectrum.  The solid curve corresponds to a
  resolution of $512^3$ and $Re=6000$, the dashed curve corresponds to a
  resolution $1024^3$ and $Re=15,000$. The $Re$ numbers are chosen as
  to ensure that $\Delta/\eta_h\approx 0.83$ is the same in both
  runs. Both simulations are essentially unresolved  at small scales.
  As a result, at $k\eta_h\gtrsim 0.1$ the solution of the numerical scheme scales differently from the inertial interval.} 
\end{figure}

The presented 
results reveal an important property of the dissipation interval of MHD turbulence. As we have seen, two scales play a role  here: the physical dissipation scale, $\eta_s$, given by~(\ref{eta_s}), and the scale, $\eta_h$, given by~(\ref{eta_h}). Suppose that the discretization scale, $\Delta$, in numerical simulations is decreased  proportionally to $\eta_s$ as the Reynolds number increases. Then, since $\eta_h$ decreases faster, at some Reynolds number we will have $\Delta\sim \eta_h$ and the simulations will become unresolved. In this case, the numerical scheme will not approximate the physical solution at scales close to $\eta_h$. If starting from this point, $\Delta$ starts to decrease proportionally to $\eta_h$, the spectral distortion will be preserved, and the approximation will not improve (cf. Fig~\ref{fig:spectra2}). However, if $\Delta$ continues to decrease slower than $\eta_h$, the approximation of the numerical scheme will continue to degrade. We therefore conclude that in order to resolve the dissipation interval of MHD turbulence, the discretization of the numerical scheme should satisfy $\Delta\ll \eta_h$, and it should decrease at least as fast as $\eta_h$ to maintain the same level of approximation.

The physical explanation of the observed phenomenon is the following. As proposed in \cite[][]{boldyrev05,boldyrev06}, the $-3/2$ scaling of the inertial interval of MHD turbulence is related to the scale-dependent dynamic alignment between magnetic and velocity fluctuations, where the alignment angle decreases with the scale approximately as $\theta(r)\propto r^{1/4}$. Numerical simulations demonstrate that such alignment is present not only in the inertial interval but also in the dissipation range as long as we do not approach the numerical cut-off too closely \cite[][]{mason_cb08,perez_etal2012}. This alignment leads to the scale-dependent weakening of the nonlinear interaction, changing the naive $-5/3$ scaling of the equation to the observed $-3/2$ scaling of the solution. 

In the unresolved intervals, however, the correlation between magnetic and velocity fluctuations gets destroyed due to the proximity to the dealiasing cut-off. A similar behavior has been documented in numerical simulations of hydrodynamic turbulence, where an abrupt cutoff in the $k$ space led to unphysical distortion of the energy spectrum \cite[][]{cichowlas_2005,frisch_etal2008,connaughton2009,grappin2010}. In the extreme case, when the explicit dissipation was totally absent, the spectral cutoff would lead to an energy pile up and mode thermalization close to the cutoff scale. At such scales any spatial correlation of fluctuations existing in the turbulent flow would be lost.

\begin{figure}
\includegraphics[width=\columnwidth]{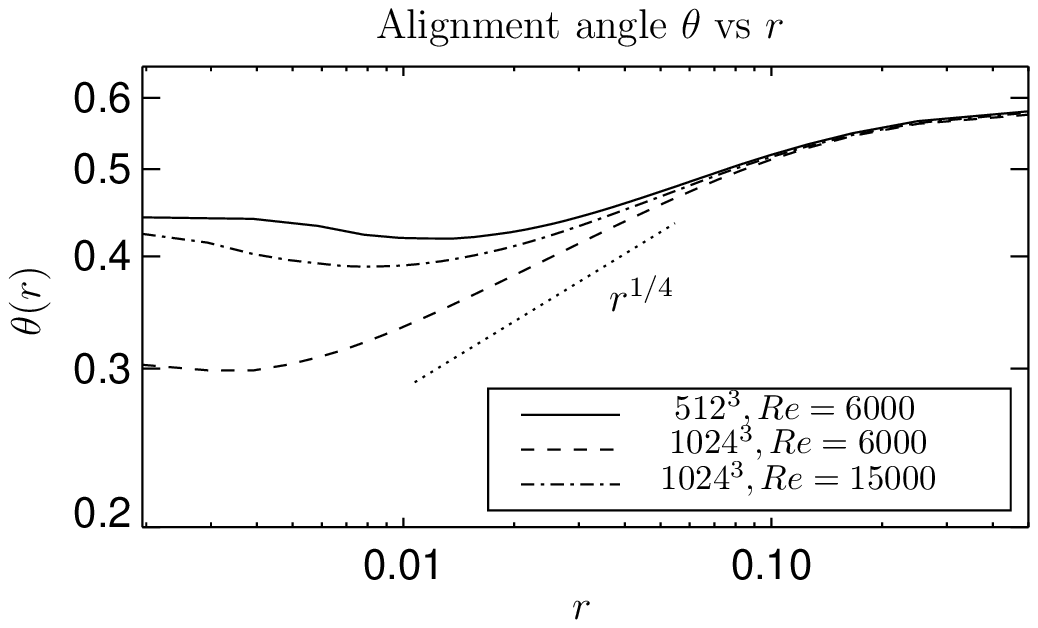}
\caption{\label{fig:spectra3}Scaling of the alignment angle $\theta(r)$ between the velocity and magnetic fluctuations measured at scale~$r$ (the numerical procedure for the measurements is described in detail in \cite[][]{perez_etal2012}). In a resolved run exhibiting the $-3/2$ scaling, the alignment $\theta(r)\sim r^{1/4}$ is well-preserved down to the scale $r\sim 0.005$ (the dashed curve). However, in the unresolved runs the alignment is already significantly spoiled at scale $r\sim 0.05$, where the curves start to deviate from the $r^{1/4}$ scaling and flatten at smaller scales (the solid and dash-dotted curves). The absence of the scale-dependent alignment leads to the $-5/3$ scaling at these scales, as observed in Figure.~\ref{fig:spectra2}.}
\end{figure}

It is reasonable to propose that a similar tendency exists in MHD turbulence when the physical dissipation is not strong enough in the vicinity of the numerical cutoff. This leads to the unphysical numerical distortion seen in Figure~\ref{fig:spectra2} and similar unphysical effects in the simulations of \citet[][]{beresnyak_11,beresnyak_12,beresnyak_14}.
In this case, the correlation properties of the fluctuations can be  affected as well. In particular, the angular alignment of magnetic and velocity fluctuations can weaken. This is indeed consistent with Figure.~\ref{fig:spectra3} which illustrates the scaling of the alignment angle between the velocity and magnetic fluctuations measured at scale~$r$ (see the details of the measurement procedure in \cite[][]{perez_etal2012}). In the  resolved run (dashed line) the alignment is well preserved down to the scale $r\sim 0.005$, corresponding to the scaling $-3/2$ observed in Figure.~\ref{fig:spectra1}. In contrast, in the unresolved runs (solid and dash-dotted lines), the alignment is already significantly spoiled at scale $r\sim 0.05$, which is consistent with the transition to the $-5/3$ scaling at smaller scales (cf. Figure.~\ref{fig:spectra2}).

In conclusion, we have demonstrated that the Kolmogorov
first self-similarity hypothesis~(\ref{espec}) does not apply to 
MHD turbulence. In contrast with the hydrodynamic case, the Fourier energy spectrum is not a universal function of the wavenumber normalized by the dissipation scale, $k\eta_s$.  This happens due to the mediation of the small-scale turbulent energy cascade by the large-scale magnetic field, which introduces the large-scale dependence in both the inertial and the dissipation intervals. We have proposed that the lack of universality has implications for numerical simulations of MHD turbulence. In particular, when the explicit physical dissipation is not strong in the vicinity of the numerical cutoff, the measured spectrum gets distorted by the numerical effects. The solution of the discrete numerical scheme in the cutoff vicinity has a Reynolds number scaling that is different from the scaling of the physical solution. This imposes stringent conditions on the numerical resolution required for correct simulations of MHD turbulence. The cutoff scale should decrease faster than the dissipation scale as the Reynolds number increases, in order for the numerical simulations to be adequately resolved.   

This work was supported by NSF/DOE grant AGS-1003451, grant NNX11AJ37G from NASA's Heliophysics Theory Program,  DoE grant DE-SC0003888, NSF grant NSF PHY11-25915, and the NSF Center for Magnetic Self-Organization in Laboratory and Astrophysical Plasmas. High Performance Computing resources were provided by the Texas Advanced Computing Center (TACC) at the University of Texas at Austin under the NSF-XSEDE Project TG-PHY110016.



\begin{thebibliography}{50}
\expandafter\ifx\csname natexlab\endcsname\relax\def\natexlab#1{#1}\fi

\bibitem[{{Beresnyak}(2011)}]{beresnyak_11}
{Beresnyak}, A. 2011, Physical Review Letters, 106, 075001

\bibitem[{{Beresnyak}(2012)}]{beresnyak_12}
---. 2012, MNRAS, 422, 3495

\bibitem[{{Beresnyak}(2014)}]{beresnyak_14}
---. 2014, \apjl, 784, L20

\bibitem[{{Bhattacharjee} \& {Ng}(2001)}]{bhattacharjee_n01}
{Bhattacharjee}, A. \& {Ng}, C.~S. 2001, Astrophys. J., 548, 318

\bibitem[{{Biskamp}(2003)}]{biskamp03}
{Biskamp}, D. 2003, {Magnetohydrodynamic Turbulence}, ed. D.~Biskamp

\bibitem[{{Boldyrev}(2005)}]{boldyrev05}
{Boldyrev}, S. 2005, Astrophys. J. Lett., 626, L37

\bibitem[{{Boldyrev}(2006)}]{boldyrev06}
---. 2006, Physical Review Letters, 96, 115002

\bibitem[{{Boldyrev} \& {Perez}(2009)}]{boldyrev_p09}
{Boldyrev}, S. \& {Perez}, J.~C. 2009, Physical Review Letters, 103, 225001

\bibitem[{{Brandenburg} \& {Nordlund}(2011)}]{brandenburg_n2011}
{Brandenburg}, A. \& {Nordlund}, {\AA}. 2011, Reports on Progress in Physics,
  74, 046901

\bibitem[{{Brandenburg} {et~al.}(2012){Brandenburg}, {Sokoloff}, \&
  {Subramanian}}]{brandenburg_etal2012}
{Brandenburg}, A., {Sokoloff}, D., \& {Subramanian}, K. 2012, \ssr, 169, 123

\bibitem[{{Chandran} {et~al.}(2011){Chandran}, {Dennis}, {Quataert}, \&
  {Bale}}]{chandran_dqb2011}
{Chandran}, B.~D.~G., {Dennis}, T.~J., {Quataert}, E., \& {Bale}, S.~D. 2011,
  \apj, 743, 197

\bibitem[{{Chandran} {et~al.}(2010){Chandran}, {Li}, {Rogers}, {Quataert}, \&
  {Germaschewski}}]{chandran_etal10}
{Chandran}, B.~D.~G., {Li}, B., {Rogers}, B.~N., {Quataert}, E., \&
  {Germaschewski}, K. 2010, Astrophys. J., 720, 503

\bibitem[{{Chen} {et~al.}(2011){Chen}, {Mallet}, {Yousef}, {Schekochihin}, \&
  {Horbury}}]{chen10}
{Chen}, C.~H.~K., {Mallet}, A., {Yousef}, T.~A., {Schekochihin}, A.~A., \&
  {Horbury}, T.~S. 2011, \mnras, 415, 3219

\bibitem[{{Cichowlas} {et~al.}(2005){Cichowlas}, {Bona{\"i}ti}, {Debbasch}, \&
  {Brachet}}]{cichowlas_2005}
{Cichowlas}, C., {Bona{\"i}ti}, P., {Debbasch}, F., \& {Brachet}, M. 2005,
  Physical Review Letters, 95, 264502

\bibitem[{{Coles} {et~al.}(2010){Coles}, {Rickett}, {Gao}, {Hobbs}, \&
  {Verbiest}}]{coles_etal2010}
{Coles}, W.~A., {Rickett}, B.~J., {Gao}, J.~J., {Hobbs}, G., \& {Verbiest},
  J.~P.~W. 2010, \apj, 717, 1206

\bibitem[{{Connaughton}(2009)}]{connaughton2009}
{Connaughton}, C. 2009, Physica D Nonlinear Phenomena, 238, 2282

\bibitem[{{Donzis} \& {Sreenivasan}(2010)}]{donzis2010}
{Donzis}, D.~A. \& {Sreenivasan}, K.~R. 2010, Journal of Fluid Mechanics, 657,
  171

\bibitem[{{Elmegreen} \& {Scalo}(2004)}]{elmegreen_s2004}
{Elmegreen}, B.~G. \& {Scalo}, J. 2004, \araa, 42, 211

\bibitem[{{Frisch} {et~al.}(2008){Frisch}, {Kurien}, {Pandit}, {Pauls}, {Ray},
  {Wirth}, \& {Zhu}}]{frisch_etal2008}
{Frisch}, U., {Kurien}, S., {Pandit}, R., {Pauls}, W., {Ray}, S.~S., {Wirth},
  A., \& {Zhu}, J.-Z. 2008, Physical Review Letters, 101, 144501

\bibitem[{{Galtier} {et~al.}(2000){Galtier}, {Nazarenko}, {Newell}, \&
  {Pouquet}}]{galtier_nnp00}
{Galtier}, S., {Nazarenko}, S.~V., {Newell}, A.~C., \& {Pouquet}, A. 2000,
  Journal of Plasma Physics, 63, 447

\bibitem[{{Goldreich} \& {Sridhar}(1995)}]{goldreich_s95}
{Goldreich}, P. \& {Sridhar}, S. 1995, Astrophys. J., 438, 763

\bibitem[{{Goldreich} \& {Sridhar}(2006)}]{goldreich_s2006}
---. 2006, \apjl, 640, L159

\bibitem[{{Gotoh}(2002)}]{gotoh2002}
{Gotoh}, T. 2002, Computer Physics Communications, 147, 530

\bibitem[{{Grappin} \& {M{\"u}ller}(2010)}]{grappin2010}
{Grappin}, R. \& {M{\"u}ller}, W.-C. 2010, \pre, 82, 026406

\bibitem[{{Haugen} {et~al.}(2004){Haugen}, {Brandenburg}, \&
  {Dobler}}]{haugen_04}
{Haugen}, N.~E., {Brandenburg}, A., \& {Dobler}, W. 2004, Physical Review E,
  70, 016308

\bibitem[{{Iroshnikov}(1963)}]{iroshnikov}
{Iroshnikov}, R.~S. 1963, Astronomicheskii Zhurnal, 40, 742

\bibitem[{{Kolekar} {et~al.}(2012){Kolekar}, {Subramanian}, \&
  {Sridhar}}]{kolekar_etal2012}
{Kolekar}, S., {Subramanian}, K., \& {Sridhar}, S. 2012, \pre, 86, 026303

\bibitem[{{Kolmogorov}(1941)}]{kolmogorov1941}
{Kolmogorov}, A. 1941, Akademiia Nauk SSSR Doklady, 30, 301

\bibitem[{{Kraichnan}(1965)}]{kraichnan65}
{Kraichnan}, R.~H. 1965, Physics of Fluids, 8, 1385

\bibitem[{{Longcope} \& {Sudan}(1994)}]{longcope_s1994}
{Longcope}, D.~W. \& {Sudan}, R.~N. 1994, \apj, 437, 491

\bibitem[{{Malyshkin} \& {Boldyrev}(2010)}]{malyshkin_b2010}
{Malyshkin}, L.~M. \& {Boldyrev}, S. 2010, Physical Review Letters, 105, 215002

\bibitem[{{Maron} \& {Goldreich}(2001)}]{maron_g01}
{Maron}, J. \& {Goldreich}, P. 2001, Astrophys. J., 554, 1175

\bibitem[{{Mason} {et~al.}(2008){Mason}, {Cattaneo}, \&
  {Boldyrev}}]{mason_cb08}
{Mason}, J., {Cattaneo}, F., \& {Boldyrev}, S. 2008, Physical Review E, 77,
  036403

\bibitem[{{Mason} {et~al.}(2012){Mason}, {Perez}, {Boldyrev}, \&
  {Cattaneo}}]{mason_etal12}
{Mason}, J., {Perez}, J., {Boldyrev}, S., \& {Cattaneo}, F. 2012, Phys.
  Plasmas, 19, 055902

\bibitem[{{Monin} \& {Iaglom}(1975)}]{monin_yaglom1975}
{Monin}, A.~S. \& {Iaglom}, A.~M. 1975, {Statistical fluid mechanics: Mechanics
  of turbulence. Volume 2 /revised and enlarged edition/}

\bibitem[{{M{\"u}ller} \& {Grappin}(2005)}]{muller_g05}
{M{\"u}ller}, W. \& {Grappin}, R. 2005, Physical Review Letters, 95, 114502

\bibitem[{{Ng} \& {Bhattacharjee}(1996)}]{ng_b96}
{Ng}, C.~S. \& {Bhattacharjee}, A. 1996, Astrophys. J., 465, 845

\bibitem[{{Ng} {et~al.}(2012){Ng}, {Lin}, \& {Bhattacharjee}}]{ng_etal2012}
{Ng}, C.~S., {Lin}, L., \& {Bhattacharjee}, A. 2012, \apj, 747, 109

\bibitem[{{Obukhov}(1941)}]{obukhov1941}
{Obukhov}, A.~M. 1941, Akademiia Nauk SSSR Doklady, 32, 22

\bibitem[{{Perez} \& {Boldyrev}(2009)}]{perez_b09}
{Perez}, J.~C. \& {Boldyrev}, S. 2009, Physical Review Letters, 102, 025003

\bibitem[{{Perez} {et~al.}(2012){Perez}, {Mason}, {Boldyrev}, \&
  {Cattaneo}}]{perez_etal2012}
{Perez}, J.~C., {Mason}, J., {Boldyrev}, S., \& {Cattaneo}, F. 2012, Physical
  Review X, 2, 041005

\bibitem[{{Podesta} \& {Bhattacharjee}(2010)}]{podesta_b10}
{Podesta}, J.~J. \& {Bhattacharjee}, A. 2010, \apj, 718, 1151

\bibitem[{{Rappazzo} {et~al.}(2008){Rappazzo}, {Velli}, {Einaudi}, \&
  {Dahlburg}}]{rappazzo_etal2008}
{Rappazzo}, A.~F., {Velli}, M., {Einaudi}, G., \& {Dahlburg}, R.~B. 2008,
  Astrophys. J., 677, 1348

\bibitem[{{Strauss}(1976)}]{strauss76}
{Strauss}, H.~R. 1976, Physics of Fluids, 19, 134

\bibitem[{{Sur} {et~al.}(2014){Sur}, {Pan}, \& {Scannapieco}}]{sur_etal2014}
{Sur}, S., {Pan}, L., \& {Scannapieco}, E. 2014, \apj, 784, 94

\bibitem[{{Terry} \& {Tangri}(2009)}]{terry09}
{Terry}, P.~W. \& {Tangri}, V. 2009, Physics of Plasmas, 16, 082305

\bibitem[{{Tobias} {et~al.}(2013){Tobias}, {Cattaneo}, \&
  {Boldyrev}}]{tobias_cb11}
{Tobias}, S.~M., {Cattaneo}, F., \& {Boldyrev}, S. 2013, {MHD Dynamos and
  Turbulence, in Ten Chapters in Turbulence}, ed. {P.A. Davidson, Y. Kaneda,
  and K.R. Sreenivasan~: Cambridge University Press, 2013}

\bibitem[{{Tsuji}(2004)}]{tsuji2004}
{Tsuji}, Y. 2004, Physics of Fluids, 16, L43

\bibitem[{{Wang} {et~al.}(2011){Wang}, {Boldyrev}, \& {Perez}}]{wang11}
{Wang}, Y., {Boldyrev}, S., \& {Perez}, J.~C. 2011, \apjl, 740, L36

\bibitem[{{Zhdankin} {et~al.}(2013){Zhdankin}, {Uzdensky}, {Perez}, \&
  {Boldyrev}}]{zhdankin_etal2013}
{Zhdankin}, V., {Uzdensky}, D.~A., {Perez}, J.~C., \& {Boldyrev}, S. 2013,
  \apj, 771, 124

\end{thebibliography}

\end{document}